\documentclass[aps,prl,reprint,superscriptaddress]{revtex4-2}
\usepackage[colorlinks=true,urlcolor=blue,citecolor=blue,linkcolor=blue,bookmarks=false,pdfstartview={FitH}]{hyperref}
\usepackage{graphicx}
\usepackage{amsmath}
\begin{document}

\title{Extremely large magnetoresistance from electron-hole compensation\\ 
in the nodal loop semimetal ZrP$_2$}

\author{J. Bannies}
\affiliation{Max Planck Institute for Chemical Physics of Solids, 01309 Dresden, Germany}
\affiliation{Quantum Matter Institute, University of British Columbia, Vancouver, British Columbia, V6T 1Z4, Canada}

\author{E. Razzoli}
\affiliation{Quantum Matter Institute, University of British Columbia, Vancouver, British Columbia, V6T 1Z4, Canada}
\affiliation{Department of Physics \& Astronomy, University of British Columbia, Vancouver, British Columbia, V6T 1Z1, Canada}

\author{M. Michiardi}
\affiliation{Quantum Matter Institute, University of British Columbia, Vancouver, British Columbia, V6T 1Z4, Canada}
\affiliation{Department of Physics \& Astronomy, University of British Columbia, Vancouver, British Columbia, V6T 1Z1, Canada}
\affiliation{Max Planck Institute for Chemical Physics of Solids, 01309 Dresden, Germany}

\author{H.-H. Kung}
\affiliation{Quantum Matter Institute, University of British Columbia, Vancouver, British Columbia, V6T 1Z4, Canada}
\affiliation{Department of Physics \& Astronomy, University of British Columbia, Vancouver, British Columbia, V6T 1Z1, Canada}
\affiliation{Max Planck Institute for Chemical Physics of Solids, 01309 Dresden, Germany}

\author{I. S. Elfimov}
\affiliation{Quantum Matter Institute, University of British Columbia, Vancouver, British Columbia, V6T 1Z4, Canada}
\affiliation{Department of Physics \& Astronomy, University of British Columbia, Vancouver, British Columbia, V6T 1Z1, Canada}

\author{M. Yao}
\affiliation{Max Planck Institute for Chemical Physics of Solids, 01309 Dresden, Germany}

\author{A.~Fedorov}
\affiliation{Leibniz Institute for Solid State and Materials Research Dresden, 01069 Dresden, Germany}
\affiliation{Helmholtz-Zentrum Berlin f\"ur Materialien und Energie, 12489 Berlin, Germany}

\author{J. Fink}
\affiliation{Leibniz Institute for Solid State and Materials Research Dresden, 01069 Dresden, Germany}
\affiliation{Max Planck Institute for Chemical Physics of Solids, 01309 Dresden, Germany}
\affiliation{Institut f\"ur Festk\"orperphysik, Technische Universit\"at Dresden, 01062 Dresden, Germany}

\author{C. Jozwiak}
\affiliation{Advanced Light Source (ALS), Berkeley, California 94720, USA}

\author{A. Bostwick}
\affiliation{Advanced Light Source (ALS), Berkeley, California 94720, USA}

\author{E. Rotenberg}
\affiliation{Advanced Light Source (ALS), Berkeley, California 94720, USA}

\author{A. Damascelli}
\affiliation{Quantum Matter Institute, University of British Columbia, Vancouver, British Columbia, V6T 1Z4, Canada}
\affiliation{Department of Physics \& Astronomy, University of British Columbia, Vancouver, British Columbia, V6T 1Z1, Canada}

\author{C. Felser}
\email[]{claudia.felser@cpfs.mpg.de}
\affiliation{Max Planck Institute for Chemical Physics of Solids, 01309 Dresden, Germany}

\begin{abstract}
Several early transition metal dipnictides have been found to host topological semimetal states and exhibit large magnetoresistance.
In this study, we use angle-resolved photoemission spectroscopy (ARPES) and magneto-transport to study the electronic properties of a new transition metal dipnictide ZrP$_2$.
We find that ZrP$_2$ exhibits an extremely large and unsaturated magnetoresistance of up to 40,000\,\% at 2\,K, which originates from an almost perfect electron-hole compensation.
Our band structure calculations further show that ZrP$_2$ hosts a topological nodal loop in proximity to the Fermi level. 
Based on the ARPES measurements, we confirm the results of our calculations and determine the surface band structure.
Our study establishes ZrP$_2$ as a new platform to investigate near-perfect electron-hole compensation and its interplay with topological band structures.

\end{abstract}

\maketitle

Topological semimetals have attracted considerable scientific interest owing to their unique electronic structure and unconventional transport properties, such as extremely large magnetoresistance~(XMR)~\cite{Kumar2017b,Wang2016c,Lou2017b,Xu2017d,Yokoi2018a,Luo2015a} and planar Hall effect~\cite{Li2018e,Li2018f}, as well as their deep connection with the high-energy particle physics~\cite{Armitage2018a}.
These unusual properties not only pave the way for future electronic and spintronic devices, but also provide a test-bed for our current models of transport theories~\cite{Hu2019a}.
The characteristic dispersion feature of most topological semimetals is a linear intersection of quasiparticle bands near the Fermi energy. In the most well studied examples, i.e. Dirac~\cite{Liu2014a,Liu2014b} and Weyl semimetals~\cite{Xu2015b,Yang2015a,Kumar2017b}, the energy bands intersect each other on a single point in \textit{k}-space with four- and two-fold degeneracies, respectively.
Nodal line semimetals, on the other hand, exhibit band crossings that extend on a one-dimensional path in \textit{k}-space~\cite{Burkov2011a}, 
which can be either an open line, a closed loop, and even loops with complex connectivity~\cite{Bzdusek2016a,Yan2017a,Chang2017d,Bi2017a}.
Examples of nodal line semimetals have been found
in ZrSiS-type materials~\cite{Nakamura2019a,Schoop2016a,Topp2016a,Takane2016a}, and early transition metal dipnictides MX$_2$ (M~=~Nb, Ta, Mo, W; X~=~P, As, Sb) in the absence of spin-orbit coupling (SOC)~\cite{Shao2019a,Xu2016a,Chen2017a,Wang2019a}.

\begin{figure*}[thb]
	\centering
	\includegraphics[width=1\linewidth]{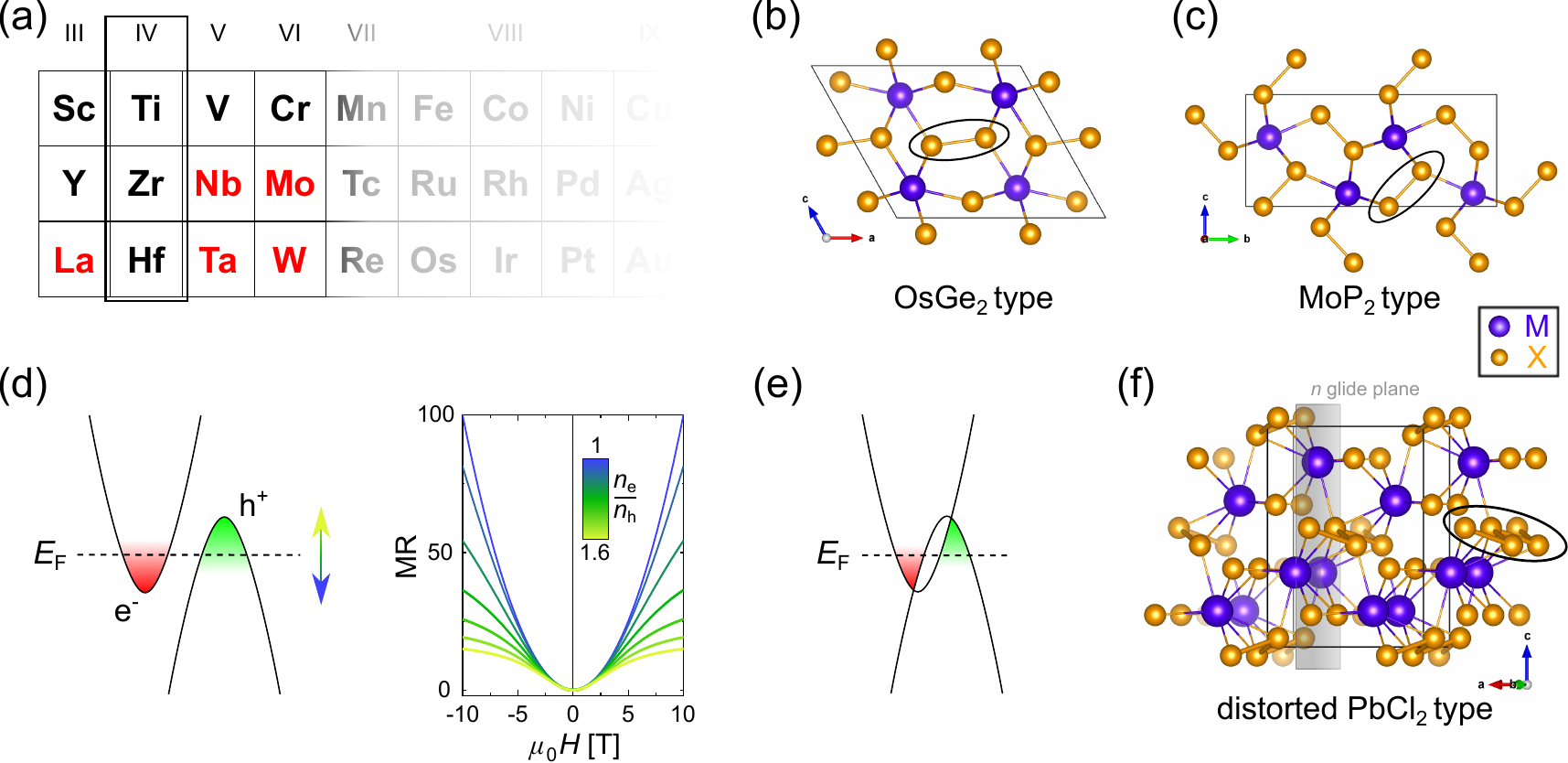}
	\caption{\label{fig:pse} Occurrence of XMR in early TMDPs.
	\textbf{(a)} Excerpt of the periodic table. Elements for which at least one TMDP was found to show XMR are marked in red. 
	\textbf{(b, c)} Common crystal structure types of early TMDPs: OsGe$_2$ type and MoP$_2$ type, both containing glides planes and two formula units per primitive unit cell. Covalent bonding motifs of anions are encircled. 
	\textbf{(d)} Schematic of a trivial semi-classical two-band model (left). Simulated MR for different carrier density ratios (right). Mobilities: $\mu_\mathrm{h}=\mu_\mathrm{e}=1$\,m$^2$/Vs. 
	\textbf{(e)} Schematic of a non-trivial compensated two-band system. 
	\textbf{(f)} Distorted PbCl$_2$ type crystal structure as found in ZrP$_2$. One of the non-symmorphic symmetry elements, the \textit{n} glide plane, is shown.}
\end{figure*}

In early transition metal dipnictides (TMDPs), non-trivial topology is often accompanied by unsaturated XMR~\cite{Yuan2016a,Wang2014a,Li2016a,Kumar2017b,Lou2017b,Du2018a}. Elements highlighted in Fig.~\ref{fig:pse}(a) all exhibit XMR in TMDPs. 
While XMR has been found for TMDPs made of group V and VI metals, as highlighted in Fig.~\ref{fig:pse}(a), no group IV TMDPs have been studied. Here, we investigate ZrP$_2$ as a case study that demonstrates the ubiquity of these properties in early TMDPs.
We propose that two factors are key for understanding the relation between topology and XMR in these compounds: non-symmorphic crystal symmetry, and partial charge transfer in the bonding.

Most of known early TMDPs crystallize either in an OsGe$_2$ type (space group~\textit{C2/m}, No.\,12) or MoP$_2$ type structure (space group~\textit{Cmc2$_1$}, No.\,36), as shown in Fig.~\ref{fig:pse}(b,c)~\cite{Hulliger1964a}. 
In the MoP$_2$ structure, the four TM sites are related by three glide planes (\textit{b}, \textit{c}, and \textit{n} glide planes) whereas in the OsGe$_2$ structure two pairs of TM sites are related by one glide plane (\textit{a} glide plane). In both cases, the conventional unit cell therefore contains four and the primitive unit cell two formula units. The non-symmorphic symmetry thus indirectly guarantees an even number of valence electrons, regardless of the atomic number of the TM. Moreover, the effective doublings of the unit cell cause band back-folding, facilitating the occurrence of band crossings such as nodal lines.

Considering that the \textit{d}~electrons of early transition metals are relatively weakly bound and easily transferable to the pnictogen~\cite{Hulliger1968a}, we would expect to see fully filled valence bands and empty conduction bands, consistent with a band insulator with no charge carriers. 
However, early TMDPs consistently show metallic conductivity in experiments~\cite{Hulliger1964a,Hulliger1968a}. 
This inconsistency is resolved by assuming that the charge transfer is incomplete, resulting in semimetallic Fermi surfaces. 
The electron- and hole-like charge carriers' density are hence naturally close to compensated, leading to an unsaturated XMR within the semiclassical approximation~(Fig.~\ref{fig:pse}(d))~\cite{Pippard1989a,Leahy2018a}. We note that the above statements should generally hold true for all early TMDPs. 
However, to the best of our knowledge, no group~IV TMDP has ever been reported to display nonsaturating XMR, leaving an open question to the generality of the above simple arguments based on non-symmorphic symmetry and incomplete charge transfer.

Here, we demonstrate that ZrP$_2$ is the first group~IV dipnictide to exhibit nonsaturating XMR. 
The electron and hole carrier densities determined from magneto-transport measurements are almost perfectly compensated, which quantitatively agrees with the observed XMR~\cite{Pippard1989a}.
The electron-hole (e-h) compensation is further confirmed by density functional theory (DFT) calculations, which also reveal the existence of a nodal loop below the Fermi level in the $k_\mathrm{x}=0$ plane.
Our band structure calculations are verified by angle-resolved photoemission spectroscopy (ARPES).

Crystals of ZrP$_2$ were grown by a two-step process. Stoichiometric mixtures of zirconium (Alfa Aesar, 99.7\,\%, foil) and red phosphorus (Alfa Aesar, 99.995\,\%, pieces) with an addition of iodine were pre-reacted at 800$^\circ$C for seven days. Subsequently,  crystals were grown by horizontal chemical vapor transport using a temperature gradient from 760\,$^\circ$C (source) to 810\,$^\circ$C (sink) with 7.5\,mg/ml iodine as transport agent. After ten days, needle-like crystals of typical dimensions 0.3x0.3x5\,mm$^3$ were obtained in the sink.
Details of the crystal characterization are given in the Supplementary Materials~\cite{SM}. The electrical transport
and Hall measurements were performed with a conventional five-probe geometry on a 9\,T PPMS (Quantum Design). The Hall data was antisymmetrized to remove contributions from the transverse magnetoresistance. 

ARPES measurements were performed at the micro-ARPES endstation of the MAESTRO beamline 7.0.2 at Advanced Light Source (ALS) and the 1$^2$ endstation of beamline UE112 PGM at BESSY.
Samples were cleaved and measured at 15\,K (20\,K) and pressures lower than $4\times 10^{-11}$\,torr ($1\times 10^{-10}$\,torr) at ALS (BESSY). 
Energy and angular resolution at both endstations are better than 20\,meV and 0.2\,$^\circ$, respectively. 

Electronic structure calculations were performed within the framework of density functional theory as implemented in the package Wien2k~\cite{Blaha2019a}. 
The generalized gradient approximation with the PBE parametrization~\cite{Perdew1996a} was used. Due to the semimetallicity of ZrP$_2$, calculations were also performed with the mBJ functional~\cite{Tran2009a}. The basis set size was set to R$_{mt}$K$_{max}$=7. For bulk calculations, the irreducible Brillouin zone (BZ) was sampled with 1440~k~points. Surface band structures were obtained from slab calculations. Slabs were constructed by stacking 5~unit cells along \textit{c} separated by 20\,Å vacuum. The BZ was sampled with a 16x30x3~k~mesh. 

\begin{figure}[ht]
	\centering
	\includegraphics[width=1\columnwidth]{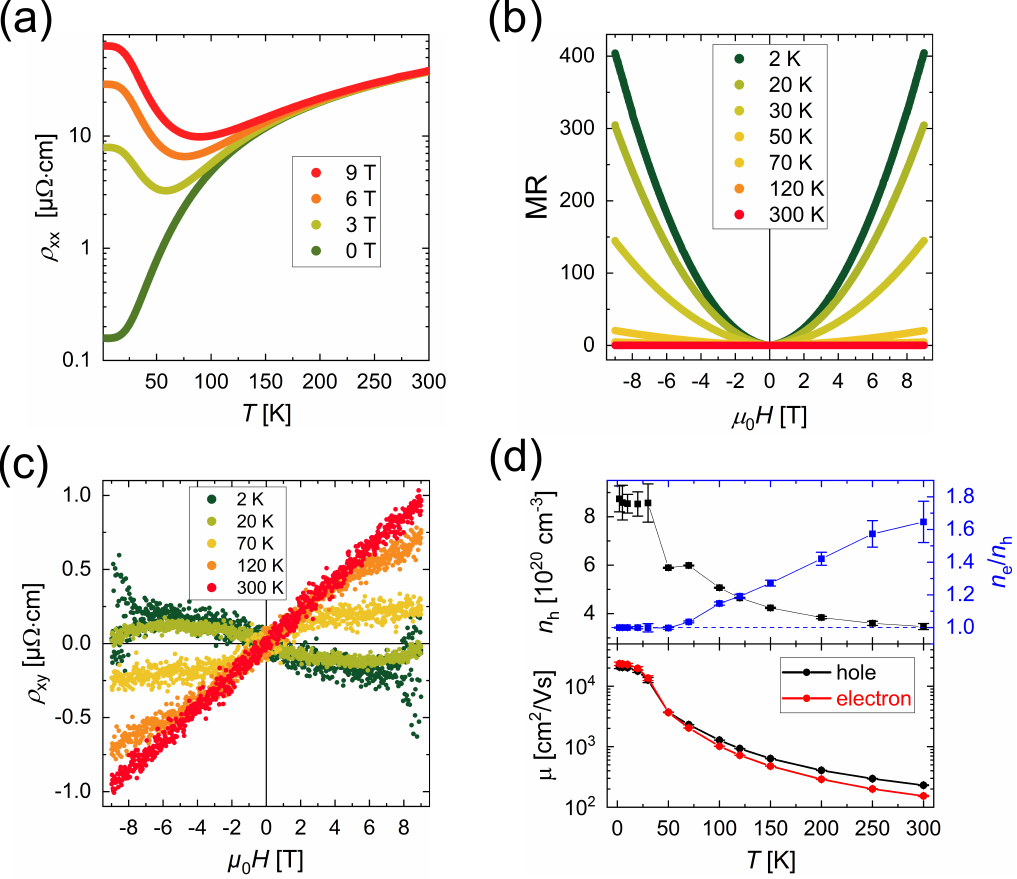}
	\caption{\label{fig:res}Electrical transport behavior of ZrP$_2$ evidencing semimetallic state. 
	\textbf{(a)} Temperature dependence of $\rho_{xx}$ in different magnetic fields with $\mathrm{I}\parallel b$. 
	\textbf{(b,c)} Magnetic field dependence of MR and $\rho_\mathrm{xy}$ at selected temperatures. 
	\textbf{(d)} Charge carrier densities and mobilities obtained by simultaneously fitting MR and $\rho_{xy}$ to the  two-band model~\cite{Pippard1989a}. Note that the large uncertainties of $n_\mathrm{h}$ compared to $n_\mathrm{e}/n_\mathrm{h}$ below 50 K originate from the fits. Depending on the initial values for the fit, values of $n_\mathrm{h}$ converged only to the window given by the uncertainties. In contrast, independent of the final value of $n_\mathrm{h}$, $n_\mathrm{e}/n_\mathrm{h}$ is always close to 1.}
\end{figure}

ZrP$_2$ crystallizes in a distorted PbCl$_2$ structure (Fig.~\ref{fig:pse}(f)) belonging to the non-symmorphic space group \textit{Pnma} (No.~62, \textit{D}$_\mathrm{2h}^{16}$)
~\cite{Huber1994a}. 
Half of the P atoms form chains with short P-P bonds (2.34\,\AA) along the \textit{b} axis indicating covalent bonding character between negatively charged P atoms (Fig.~\ref{fig:pse}(f)). These chains provide a preferred direction of growth. Indeed, crystals obtained from chemical vapor transport exhibit needle-like shape with the crystallographic \textit{b} axis oriented along the needle axis. 
The lattice parameters $a=6.4933(2)$\,\AA, $b=3.5120(1)$\,\AA, and $c=8.7431(3)$\,\AA~determined from powder XRD agree with previously reported values~\cite{Huber1994a}.

The resistivity measurements reveal metallic behavior with a low residual resistivity of 0.18\,$\mu\Omega\cdot\mathrm{cm}$ at 2\,K (Fig.~\ref{fig:res}(a)). The high residual resistivity ratio of 238 is indication of a good quality of the crystals. Notably,
a field-induced metal-to-insulator-like transition is observed above 3\,T and below 50\,K~(Fig.~\ref{fig:res}(a)), similar to other topological semimetals~\cite{Yuan2016a,Wang2014a,Li2016a,Kumar2017b}.
As shown in Fig.~\ref{fig:res}(b), the MR, which is defined as $\mathrm{MR}=\rho(B)/\rho(B=0)-1$, remains unsaturated in fields up to 9\,T and reaches 40,000\,\% at 2\,K and 9\,T. 
The field dependence of the MR is close to parabolic and can be fitted to a power law $\mathrm{MR}=a+b\cdot B^c$ with $c=1.92$.

To investigate the origin of the large and unsaturated MR in detail, we performed Hall measurements
where the resistivity $\rho_{xy}(B)$ is perpendicular to both the field and electric current. 
The $\rho_{xy}(B)$ curves (Fig.~\ref{fig:res}(c)) feature a distinct non-linearity that can be fitted with a semiclassical two-band model~\cite{Pippard1989a}. 
We note that the noise level of these curves is inherent to ZrP$_2$ due to the narrow width of the crystals.
By simultaneously fitting MR and Hall resistivity to the two-band model, we obtained carrier densities $n_\mathrm{e,h}$ and mobilities $\mu_\mathrm{e,h}$ as shown in Fig.~\ref{fig:res}(d). At room temperature, electrons are the dominant charge carriers. Upon cooling, the ratio $n_\mathrm{e}/n_\mathrm{h}$ decreases and the charge carriers become almost completely compensated below 50\,K, with $n_\mathrm{e}/n_\mathrm{h}=0.999(2)$ at 2\,K. The mobilities of electrons and holes also show similar values for all temperatures and reach high values up to $2.3\times10^4$\,cm$^2$/Vs at 2\,K, which further confirms the good crystal quality. Together, our magneto-transport measurements strongly suggest that e-h compensation is the origin for the unsaturated XMR in the semimetal ZrP$_2$.  

\begin{figure*}[ht]
	\centering
	\includegraphics[width=1.0\linewidth]{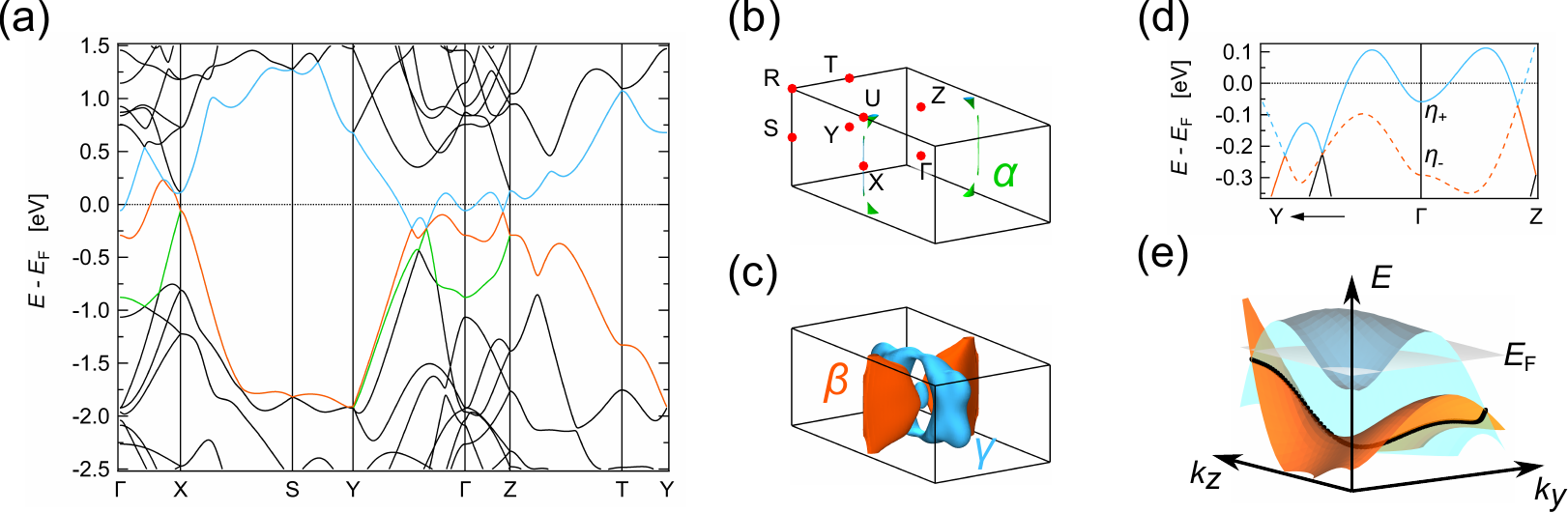}
	\caption{\label{fig:dft}Electronic structure of ZrP$_2$ from DFT calculations without SOC. 
	\textbf{(a)} Band structure with the three bands crossing $E_\mathrm{F}$ colored. 
	\textbf{(b,c)} 3D Fermi surface of ZrP$_2$ exhibiting two hole pockets ($\alpha$ and $\beta$) and an electron pocket ($\gamma$). Color coding according to panel a. 
	\textbf{(d)} Close-up of the low-energy band structure in the $k_\mathrm{x}=0$ plane. Eigenvalues $\eta_\pm$ with respect to the \textit{n}-glide plane are indicated (closed line $\eta_+$, dashed $\eta_-$). 
	\textbf{(e)} 3D visualization $E(k_\mathrm{y},k_\mathrm{z})$ of the $\beta$ and $\gamma$ bands forming the nodal loop (black line) in the $k_\mathrm{x}=0$ plane. Only one quadrant of the $k_\mathrm{x}=0$ plane is shown.}
\end{figure*}

In support for these findings, we performed DFT calculations and the resulting band structure is presented in Fig.~\ref{fig:dft}(a). 
ZrP$_2$ shows semimetallicity with both electron and hole-like bands crossing the Fermi level. 
The corresponding Fermi surface is formed by two hole-like bands, $\alpha$ and $\beta$ sheets in Fig.~\ref{fig:dft}(b,c), and one electron-like band $\gamma$, see Fig.~\ref{fig:dft}(c). 
However, the volume of the $\alpha$ hole pocket is very small compared to the other pockets and can be neglected without loosing soundness of the model.
The description by an effective two-band model remains valid and in accordance with transport experiments. The calculated ratio of the carrier densities $n_\mathrm{e}/n_\mathrm{h}$ equals 1.02, demonstrating almost perfect e-h compensation in line with the electrical transport experiments.

A close look to the low energy band structure reveals a nodal loop in the $k_\mathrm{x}=0$ plane, with energy slightly below the Fermi level, as shown in Fig.~\ref{fig:dft}(e). 
The stability of the nodal loop against opening a gap is guaranteed by the 
\textit{n}-glide plane (Fig.~\ref{fig:pse}(f)), i.e., the bands have eigenvalues of opposite sign with respect to the glide reflection symmetry, as indicated in Fig~\ref{fig:dft}(d).
Upon inclusion of SOC, the nodal loop opens a small gap, which varies between 7 to 20\,meV.

\begin{figure*}[ht]
	\centering
	\includegraphics[width=1.0\linewidth]{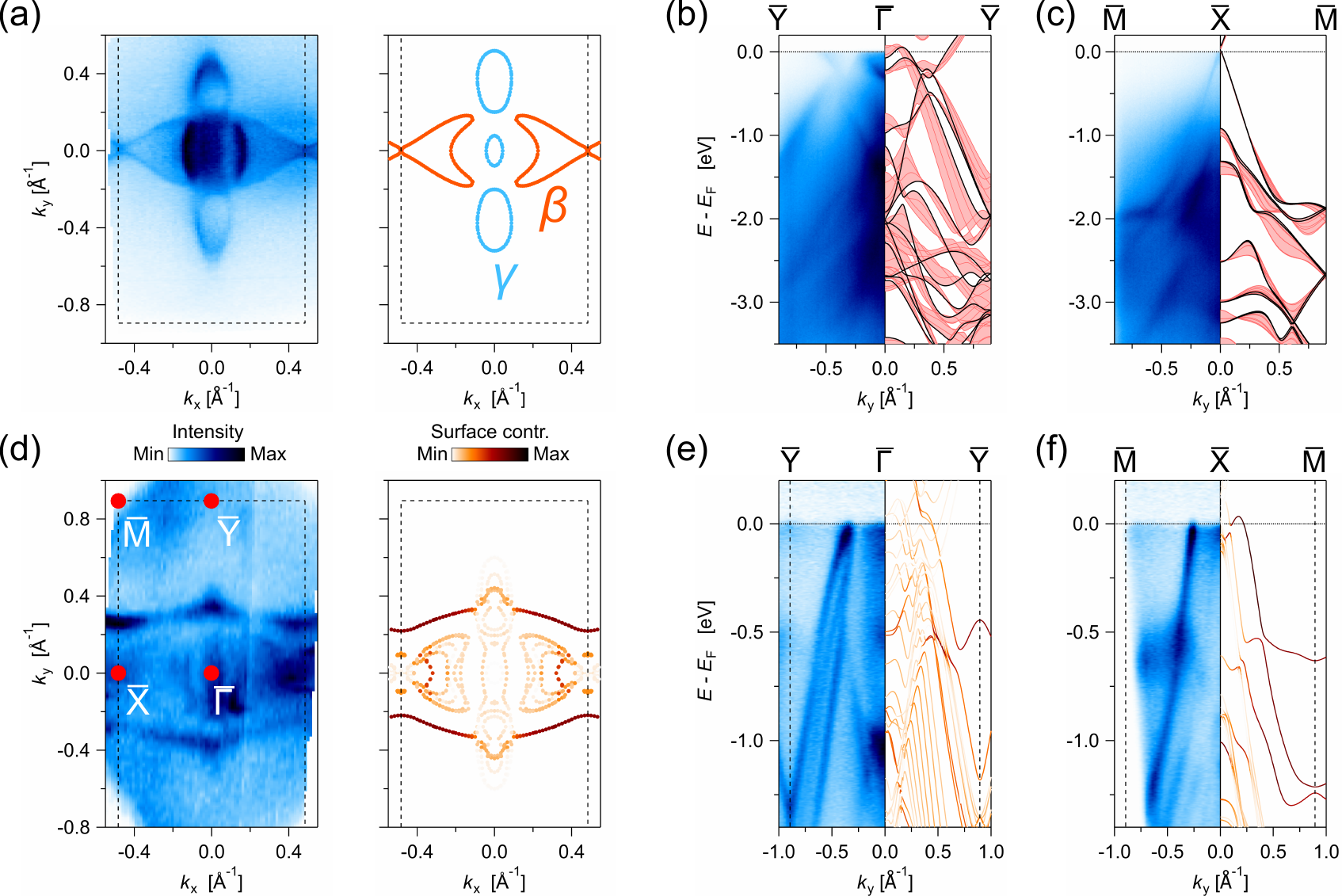}
	\caption{\label{fig:arpes}Bulk and surface band structure of ZrP$_2$ measured by ARPES (both $\mathrm{h}\nu=40$\,eV, linear horizontal polarization) and comparison with DFT calculations. 
	\textbf{(a)} Bulk Fermi surface from ARPES (left) and calculations (right) both at $k_\mathrm{z}=0$. 
	\textbf{(b,c)} Cuts along $\mathrm{\bar{Y}}$-$\bar{\Gamma}$-$\mathrm{\bar{Y}}$ and $\mathrm{\bar{M}}$-$\mathrm{\bar{X}}$-$\mathrm{\bar{M}}$ showing bulk bands, respectively. Calculations show the projected band structures for $k_\mathrm{z}\in[0,0.6\pi/c]$, where bands with $k_\mathrm{z}=0$ are highlighted in black. 
	\textbf{(d-f)} Data acquired at ALS, where the small spot size allowed us to focus on an area with a single termination, emphasizing the surface band structures~\cite{SM}. Note that the calculated surface band structure was shifted by 50\,meV with respect to the theoretical Fermi level to match the experimental dispersion. Experimental Fermi surfaces were integrated within $\pm$10\,meV of the Fermi level.}
\end{figure*}

In order to get a conclusive picture on the electronic structure of ZrP$_2$, we employed ARPES as a tool to directly map the band structure. The experimental bulk Fermi surface shown in Fig.~\ref{fig:arpes}(a) was measured with 40\,eV photons at BESSY.
This photon energy was chosen to correspond to $k_\mathrm{z}=0$, as determined by photon-energy dependent measurements (see SM~\cite{SM}). The data confirms the presence of both the $\beta$ hole pocket and the $\gamma$ electron pocket, in good agreement with theory and transport experiments. The $\alpha$ hole pocket is too small to be identified, given the resolution of our ARPES data. 

We further confirm the agreement between theory and ARPES by investigating the high-symmetry paths $\overline{\mathrm{Y}}$-$\overline{\Gamma}$-$\overline{\mathrm{Y}}$ and $\overline{\mathrm{M}}$-$\overline{\mathrm{X}}$-$\overline{\mathrm{M}}$ in a large binding energy window, as illustrated in~Fig.~\ref{fig:arpes}(b,c). 
At the low photon energy used in our experiment, $k_\mathrm{z}$ is ill-defined, as a consequence of the short mean free path of the photoemitted electron, and the relatively small size of the Brillouin zone along~$k_\mathrm{z}$~\cite{Strocov2003a}. We find that the uncertainty of $k_\mathrm{z}$ is ca.~0.6\,$\pi/c$ (see SM~\cite{SM} for details).
Therefore, we compare the spectra with the projected bulk band structure within the $k_\mathrm{z}\in[0,0.6\pi/c]$ range, and find good agreement in the entire binding energy range. 
We note that the band crossing point of the nodal loop along $\overline{\mathrm{Y}}$-$\overline{\Gamma}$-$\overline{\mathrm{Y}}$ could not be identified owing to two possibly concomitant reasons: i) at any measured photon energy only one of the two crossing bands is seen, due to unfavorable photoemission cross section~\cite{Damascelli2004a}; ii) broadening from surface disorder and $k_\mathrm{z}$ integration smears out the band crossing.

In addition to the bulk band structure, we have studied the surface states of ZrP$_2$. Generally speaking, the appearance of surface states in ARPES depends on experimental conditions such as the quality and topography of the cleaved surface. This is of particular importance for 3D materials like ZrP$_2$, where cleaving may reveal different surface terminations~\cite{SM}. Here, a small beam spot size can help to probe a single termination in the ARPES experiment. For this reason, we conducted additional ARPES studies using microARPES, the results of which are shown in Fig.~\ref{fig:arpes}(d-f). The spectra are dominated by surface states, in contrast to the data shown in Fig.~\ref{fig:arpes}(a-c).
The projected bulk Fermi surface is enclosed by intense surface states, which disperse over more than 1\,eV. 
Along $\overline{\mathrm{M}}$-$\overline{\mathrm{X}}$-$\overline{\mathrm{M}}$, the surface states are well separated from the projected bulk band structure, whereas they partly hybridize with the bulk along $\overline{\mathrm{Y}}$-$\overline{\Gamma}$-$\overline{\mathrm{Y}}$. 
Altogether, we find excellent agreement between the measured spectra and the slab calculations
considering a surface termination that exposes the phosphorus chains without breaking them~\cite{SM}.

The presence of both covalent and ionic bonding features in ZrP$_2$, namely the phosphorus chains along the \textit{b}\nobreakdash-axis, indicates a significant degree of charge transfer from zirconium to phosphorus.
However, in contrast to the insulating behaviour favored by complete charge transfer, our transport and spectroscopic measurements clearly evidence a charge compensated semimetallic ground state in ZrP$_2$. 
The e-h compensation at low temperatures is further confirmed by band structure calculations 
and accounts for the unsaturated XMR~\cite{Pippard1989a}. 
In this regard, ZrP$_2$ resembles the behavior of other early TMDPs such as NbAs$_2$~\cite{Wang2016c,Yokoi2018a} and WP$_2$~\cite{Kumar2017b,Razzoli2018a,Schoenemann2017a}. 
With the finding of XMR in ZrP$_2$, we are able to establish XMR as a general feature in early TMDPs.
The underlying reason for this ubiquity is the combination of non-symmorphic symmetry and incomplete charge transfer. 

Strictly speaking, semimetallicity alone is not sufficient to explain the unsaturated XMR in all of the reported materials.
Within the semi-classical two-band model, unsaturated MR is only possible if the e-h densities are exactly compensated~\cite{Pippard1989a}.
In the more realistic cases, XMR is associated with the combination of moderate to perfect e-h compensation and sufficiently high carrier mobility.
Such conditions are commonly fulfilled in topological semimetals, where the band crossings close to the Fermi energy give rise to very high carrier mobilities (Fig.~\ref{fig:pse}(e)). 
It has also been shown in several cases that an imbalanced e-h compensation can be offset by the very high carrier mobility, such that XMR is still observed~\cite{Jiang2018a,Liang2015a,Tafti2015a}.
Many of the early TMDPs were predicted and, in some cases, verified to host topologically non-trivial phases~\cite{Kumar2017b,Razzoli2018a,Xu2016a,Gresch2017a,Wang2019a}. 
However, the coexistence of e-h compensation and non-trivial topology naturally complicates the disentanglement of their influence on transport, which has caused difficulties interpreting the origin of the XMR~\cite{Wang2014a}. 
Based on our DFT calculations and ARPES measurements, we argue that ZrP$_2$ is a clean case where the unsaturated XMR can be fully accounted for by trivial e-h compensation.
Contributions from the non-trivial nodal loop can be excluded as it lies ca. 70-230\,meV below the Fermi level. Additionally, open-orbit Fermi surface topology as in MoAs$_2$~\cite{Lou2017b,Singha2018a} can be ruled out as the origin. 
A similar situation was found for the type-II Weyl semimetal WP$_2$ for which the XMR was attributed to e-h compensation despite the existence of Weyl nodes~\cite{Razzoli2018a}.

The direct observation of the nodal loop in the $k_\mathrm{x}=0$ plane by ARPES was prevented by experimental limitations. Nevertheless, the excellent agreement between our calculations and the ARPES spectra provides a strong support for its presence in the low energy band structure of ZrP$_{2}$. Inclusion of soft X-ray ARPES in future investigations might be a powerful tool to 
reduce the $k_\mathrm{z}$~broadening and resolve the nodal loop along $k_\mathrm{z}$.

One might wonder about the topological nature of the surface states seen by ARPES and DFT. Nodal line semimetals are expected to host the so called drumhead topological surface states~\cite{Burkov2011a}. However, the presence of a drumhead surface state is only guaranteed on the (100) surface of ZrP$_2$, whereas ZrP$_2$ naturally cleaves on the (001) surface.
Together with the strongly termination dependent connectivity of the (001) surface states found in our calculations, we conclude that the observed surface states are trivial in nature. This contrasts a recent report on HfP$_2$~\cite{Sims2020a}, a heavier homologue of ZrP$_2$, where similar (001) surface states were interpreted as topological. Minor differences in the bulk band structures of ZrP$_2$ and HfP$_2$ may account for this contrast.

\begin{figure}[ht]
	\centering
	\includegraphics[width=1\columnwidth]{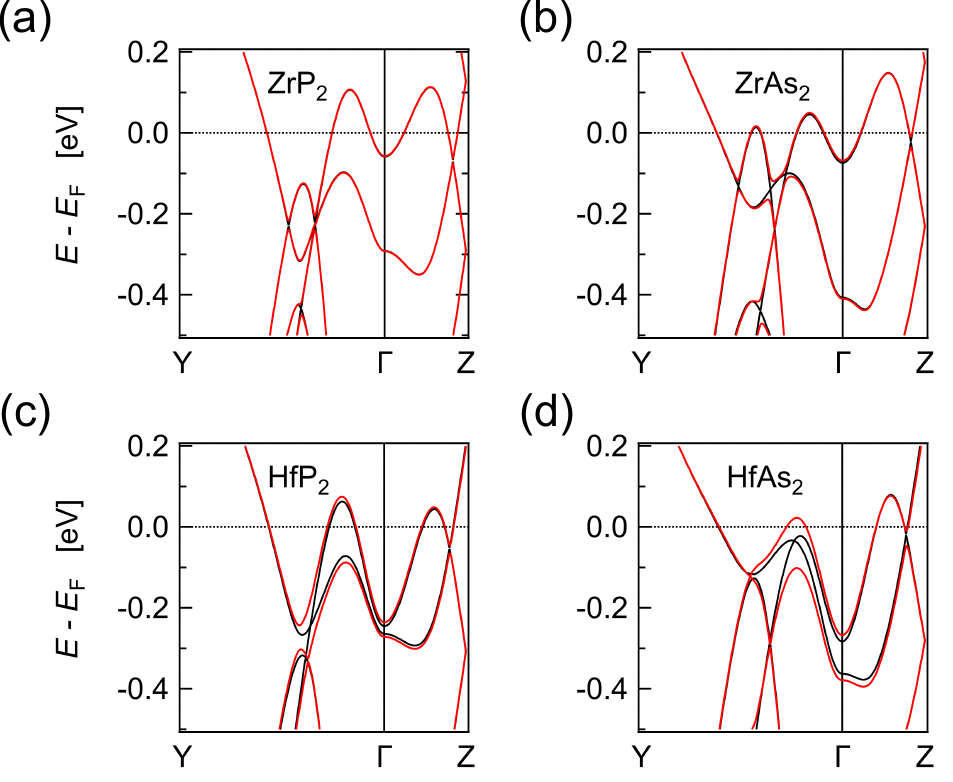}
	\caption{\label{fig:fam}Gap opening along the nodal loop through increased SOC. Band structures along Y-$\Gamma$-Z for \textbf{(a)} ZrP$_2$, \textbf{(b)} ZrAs$_2$, \textbf{(c)} HfP$_2$, and \textbf{(d)} HfAs$_2$. Black lines: without SOC, red lines: with SOC. Experimentally determined crystal structures were used for calculations on ZrP$_2$~\cite{Huber1994a} and ZrAs$_2$~\cite{Blanchard2010a}. For HfP$_2$ and HfAs$_2$, the crystal structures of ZrP$_2$ and ZrAs$_2$ were used, respectively.}
\end{figure}

The structural similarity between ZrP$_2$ and HfP$_2$ can be further exploited. ZrP$_2$ has three heavier homologues: ZrAs$_2$, HfP$_2$ and HfAs$_2$ belonging to the same structure type~\cite{Hulliger1968a,Jeitschko1962b}. The band structures are therefore very similar to ZrP$_2$. In particular, all these materials feature a nodal loop in the $k_\mathrm{x}=0$ plane (see Fig.~\ref{fig:fam}). The strength of SOC, however, increases significantly between ZrP$_2$ (lightest) and HfAs$_2$ (heaviest). 
Accordingly, the gap opened along the nodal loop by SOC increases from 20\,meV in ZrP$_2$ to 120\,meV in HfAs$_2$. The ZrP$_2$ family of materials thus provides a good platform to study the evolution of nodal loops with increasing SOC, an aspect of nodal loops that has received little attention so far.
Furthermore, substitution of the pnictide with Ge, Si and Se has been demonstrated~\cite{Blanchard2010a,Gaultois2010a,Ishida2016a}, which opens the possibility of shifting the chemical potential by doping and access the topological crossings.
We note that the corresponding diantimonides ZrSb$_2$ and HfSb$_2$ exist as well but have been shown to exhibit rather metallic behavior~\cite{Sun2017a}, presumably due to a lesser degree of charge transfer from zirconium/hafnium to antimony. In addition, the diantimonides crystallize in a different structure and space group than the diphosphides and diarsenides~\cite{Kjekshus1972a}. Hence, we cannot directly compare their band structures.

In summary, we demonstrated that ZrP$_2$ is an e-h compensated nodal loop semimetal. By applying combined experimental and theoretical methods, we derived a consistent picture of the electronic structure and the resulting properties. The extremely large and unsaturated MR of 40,000\,\% at 2\,K and 9\,T is explained by trivial e-h compensation. Our band structure calculations, showing a nodal loop in the $k_\mathrm{x}=0$ plane, are confirmed by ARPES experiments. However, contributions from non-trivial topology can be excluded as the origin of the XMR.
Our findings establish unsaturated XMR as a general consequence of non-symmorphic symmetry and incomplete charge transfer in early TMDPs. The group IV TMDPs, including ZrP$_2$, stand out of this class of materials because of their tunability both in terms of strength of SOC and possibility to shift the chemical potential by doping. They thus provide an ideal platform for future investigations of nodal lines.

We thank R. Koban and W. Schnelle for assistance with the magnetotransport measurements. This research was undertaken thanks in part to funding from the ERC Advanced Grant No. (742068) `TOPMAT'; the Max Planck-UBC-UTokyo Centre for Quantum Materials and the Canada First Research Excellence Fund, Quantum Materials and Future Technologies Program, in addition to the Killam, Alfred P. Sloan, and Natural Sciences and Engineering Research Council of Canada’s (NSERC’s) Steacie Memorial Fellowships (A.D.); the Alexander von Humboldt Fellowship (A.D.); the Canada Research Chairs Program (A.D.); NSERC, Canada Foundation for Innovation (CFI); British Columbia Knowledge Development Fund (BCKDF); and the CIFAR Quantum Materials Program.
This research also used resources of the Advanced Light Source, which is a DOE Office of Science User Facility under Contract No. DE-AC02-05CH11231.

\end{document}